\documentclass[12pt,tightenlines,
    onecolumn,
    preprintnumbers,
    amsmath,
    amssymb,
    prd,
    showpacs,
    showkeys
]{revtex4}
\usepackage{colordvi}
\input colordvi
\usepackage{color}

\usepackage{graphicx}%
\usepackage{amsmath}
\usepackage{amssymb}
\usepackage{bm}
\usepackage{enumerate}

\preprint{CQUeST-2008-0202}

\newcommand{\del}{\partial}
\newcommand{\nn}{\nonumber}

\newcommand{\beq}{\begin{equation}}
\newcommand{\eeq}{\end{equation}}
\newcommand{\beqa}{\begin{eqnarray}}
\newcommand{\eeqa}{\end{eqnarray}}
\newcommand{\bseq}{\begin{subequations}}
\newcommand{\eseq}{\end{subequations}}

\newcommand{\fr}{\frac}
\newcommand{\bg}{\textbf{g}}

\newcommand{\cF}{\mathcal{F}}

\newcommand{\cU}{\mathcal{U}}

\newcommand{\cFk}{\mathcal{F_\kappa}}
\begin{document}
\title{Differential structure on the $\kappa$-Minkowski spacetime from twist}

\author{Hyeong-Chan Kim}
\email{hyeongchan@sogang.ac.kr}
\author{Youngone Lee
}
\email{ youngone@yonsei.ac.kr}
\affiliation{Center for Quantum Space Time, Sogang University,
Seoul 120-749, Republic of Korea}
\author{Chaiho Rim}
\email{rim@chonbuk.ac.kr}
\affiliation{Department of Physics and Research Institute
of Physics and Chemistry, Chonbuk National University, Jeonju 561-756, Korea}
\author{Jae Hyung Yee}
\email{jhyee@yonsei.ac.kr}
\affiliation{Department of Physics, Yonsei University,
Seoul 120-749, Republic of Korea}
\bigskip
\begin{abstract}
\bigskip
We study four dimensional $\kappa$-Minkowski spacetime constructed by the twist deformation of $U(igl(4,R))$.
We demonstrate that the differential structure of such twist-deformed $\kappa$-Minkowski spacetime is closed in four dimensions contrary to the construction of $\kappa$-Poincar\'{e} bicovariant calculus which needs an extra fifth dimension.
Our construction holds in arbitrary dimensional spacetimes.

\end{abstract}
\maketitle

\subsubsection{Introduction}
There has been much interest in recent years in a possible role of deformation of spacetime symmetry in describing Planck scale physics. In particular,
initiated by the $\kappa$-deformed Poincar\'{e} algebra~\cite{kappaP}
the $\kappa$-Minkowski spacetime~\cite{majid,zakr}
satisfying
\begin{eqnarray}\label{kappa}
[x^0, x^i]= \frac{i}{\kappa} x^i, \quad [x^i,x^j]=0,
\end{eqnarray}
has attracted much attention in explaining cosmic observational data,
since the deformation preserves the rotational symmetry in space.
The differential structure of the $\kappa$-Minkowski spacetime has been
constructed in~\cite{kappa-diff} and based on this differential structure,
the scalar field theory has been formulated~\cite{kosinski,KRY-a,int-scalar}.
Similar field theoretic approach is given in~\cite{arzano1,Meljanac} using the coproduct and star product as Lie-algebraic noncommutative spacetime.
It was shown that the differential structure requires that the momentum space
corresponding to the $\kappa$-Minkowski spacetime becomes a de-Sitter section in
five-dimensional flat space.
The $\kappa$-deformation was extended to the curved space with
$\kappa$-Robertson-Walker metric and was applied to the cosmic microwave
background radiation in~\cite{kim}.
The physical effects of the $\kappa$-deformation has been investigated on the
Unruh effect~\cite{KRY-a}, black-body radiation~\cite{KRY-b} and Casimir
effect~\cite{Casimir}.
The Fock space and its symmetries~\cite{amelino,arzano}, 
$\kappa$-deformed statistics of particles~\cite{das,other2}, 
and interpretation of the $\kappa$-Minkowski spacetime in 
terms of exotic oscillator \cite{Ghosh} were also studied.

Recently,simpler realization of the $\kappa$-Minkowski
spacetime by the use of twisting procedure have been sought by 
several authors~\cite{lightlike,Bu,other,twist:other2,Ballest}.
It happens that only the case of the light-cone 
$\kappa$-deformation the deformed Poincar\'e algebra can be 
described by standard twist  (see eg.~\cite{lightlike}).

By embedding an abelian twist in $IGL(4,R)$
whose symmetry is larger than the Poincar\'e,
the realization for the time-like $\kappa$-deformation was 
first constructed in Ref.~\cite{Bu} and then by~\cite{other}.
Some physical properties of analogous twist realization of 
$\kappa$-Minkowski spacetime were discussed  
recently~\cite{twist:other2}.
This approach can be seen as an alternative to the 
$\kappa$-like deformation of the quantum 
Weyl and conformal algebra~\cite{Ballest}, 
which is obtained by using the Jordanian 
twist~\cite{jordantwist}.
One may even consider the chains of twists for classical Lie algebras~\cite{kulish}.

In this letter, we will construct
the $\kappa$-Minkowski spacetime and
its differential structure using the twisted
universal enveloping Hopf algebra
of the inhomogeneous
general linear group in (3+1)-dimensions.
In section 2, the $\kappa$-Minkowski spacetime from twist
is reviewed and in section 3,
its differential structure is constructed.

\subsubsection{Review on the $\kappa$-Minkowski spacetime from twist}

Twisting the Hopf algebra of the universal enveloping algebra of $igl(4,R)$ is considered in~\cite{Bu,other}.
The group of inhomogeneous linear coordinate transformations is composed of the product of the general linear transformations and the spacetime translations.
The inhomogeneous general linear algebra in (3+1)-dimensional flat spacetime $\bg=igl(4,R)$ is composed of $20$ generators $\{P_a,M^a_{~b}\}$ $(a,b=0,1,2,3)$ where $P_a$ represents the spacetime translation and $M^a_{~b}$ homogeneous one including the boost generator, rotation and dilation.
The generators satisfy the commutation relations,
\begin{eqnarray}
\label{iglbasis}
\left[P_a, P_b\right] &=&0, \quad
\left[ M^a_{~b}, P_c \right] = i\delta^a_{~c}\cdot P_b, \nn \\
\left[M^a_{~b} ,M^c_{~d}\right] &=&
    i\left(\delta^a_{~d}\cdot M^c_{~b}-\delta^c_{~b}
    \cdot M^a_{~d}\right).
\end{eqnarray}
The universal enveloping Hopf algebra
$\cU(\bg)$ can be constructed starting
from the base elements $\{1, P_a, M^a_{~b}\}$
and coproduct $\Delta Y=1\otimes Y+Y\otimes 1$
with $Y\in \{P_a, ~M^a_{~b}\}$.
The operators representing energy $E$
and spatial dilatation $D$ is defined by
\begin{eqnarray}\label{E}
 E= P_0,\quad \quad
 D=\sum_{i=1}^3 M^i_{~i}.
\end{eqnarray}
Note that the two generators $D$ and $E$
commutes with each other, $[D, E]=0$.
In Ref.~\cite{Bu}, an Abelian twist element $\cFk$,
\begin{eqnarray}
\label{kelement}
{\cal F_\kappa}=
\exp\left[\frac{i}{\kappa}
\Big(\alpha E\otimes D- (1-\alpha) D\otimes E \Big)\right],
\end{eqnarray}
is shown to
generate the $\kappa$-Minkowski spacetime
with the twisted Hopf algebra $\cU_\kappa(\bg)$.
$\alpha$ is a constant
chosen as $\alpha=1/2$ in this letter
which corresponds to the symmetric ordering
of the exponential kernel function in the conventional
$\kappa$-Minkowski spacetime formulation.
Other choice of $\alpha $ represents a different ordering.

Co-unit and antipode are not twisted
$\epsilon_{\cF}= \epsilon$ and $S_\cF=S$,
but coproduct is twisted as
\begin{equation}
\label{def-coproduct}
\Delta_\kappa(Y)=\cF_\kappa\cdot \Delta Y\cdot \cF_\kappa^{-1}=\sum_{i}Y_{(1)i}\otimes Y_{(2)i}\equiv Y_{(1)} \otimes Y_{(2)} .
\end{equation}
Explicitly, ($i,j=1,2,3$)
\begin{eqnarray} \label{iglcoproduct}
\Delta_\kappa(Z) &=& Z \otimes 1+ 1\otimes Z\,, \quad
		Z\in \{ E, D, M^0_{~0}, M^i_{~j}\},  \\
\Delta_\kappa(P_i) &=& P_i\otimes
    e^{E/(2\kappa)}  + e^{-E/(2\kappa)}\otimes P_i, \nn\\
\Delta_\kappa(M^i_{~0}) &=& M^i_{~0}\otimes
    e^{-E/(2\kappa)}  + e^{E/(2\kappa)}\otimes M^i_{~0}, \nn \\
\Delta_\kappa(M^0_{~i}) &=& M^0_{~i}\otimes
    e^{E/(2\kappa)}  + e^{-E/(2\kappa)}\otimes M^0_{~i}+
    \frac{1}{2\kappa}\left(P_i\otimes D e^{E/(2\kappa)}-e^{-E/(2\kappa)} D\otimes P_i\right). \nn
\end{eqnarray}
It is noted that the twisted Hopf algebra
is different from that of the conventional
$\kappa$-Poincar\'e algebra in two aspects.
First, the algebraic part is nothing but those of
the un-deformed inhomogeneous general linear group
(\ref{iglbasis})
rather than that of the deformed Poincar\'{e}.
Second, the co-algebra structure is enlarged
due to the bigger symmetry $igl(4)$
and its co-product is deformed as (\ref{iglcoproduct}).

\subsubsection{Differential structure}

The inhomogeneous general linear group $IGL(4,R)$
acts on the coordinate space $\{x^a\}$ and
the twisted-coproduct of the generator $Y$
acts on the tensor product
space of $\{ x^a \otimes x^b\}$.  Thus,
one can define the $\ast$-product of the coordinate vectors
$ x^a$ in terms of the twist action on the coordinates.
Explicitly,
\begin{eqnarray}
x^a\ast x^b\equiv\ast[x^a \otimes x^b] &=&
\cdot \left[ \cF_\kappa^{-1}
\triangleright (x^a \otimes x^b) \right] .
\end{eqnarray}
This results in the noncommutative commutation
relation of the coordinates
\[
~[x^0, x^j]_\kappa \equiv x^0 \ast x^j-x^j\ast x^0=\frac{i}{\kappa}x^j,
~~~~~ [x_i,x_j]_\kappa=0,
\]
which reproduces the commutation relation~(\ref{kappa}).

To understand the differential structure,
one has to incorporate the (co-)tangent space
and investigate the action of $IGL(4,R)$ on the space.
Suppose that one constructs
a set of basis vectors of a coordinate system
$CS=\{ e_a| a=0,1,2,3 \}$
of the four dimensional vector space $V_4$
which are not necessarily ortho-normal.
One naturally demands that
the homogeneous transformation $\Lambda$ acts
on the coordinates $x^a$, the dual-basis of the
coordinate system $e^a$, and a function $f$ as
\begin{eqnarray}\label{Lambda:f}
\Lambda &:&\left\{\begin{tabular}{clll}
    $ x^a$ & $ \to $& $x^{a'} ;$
    	& $ {x}^{a'}= x^b {\Lambda}_b^{~a'}$, \\
    $ e^a$ & $\rightarrow $& $e^{a'}$;
    	& 	$ e^{a'} = e^b {\Lambda}_b^{~a'}$, \\
  $ f$& $\rightarrow $&$ f'$ ; &$ f'(x') = f(x)= f(x^{b'} (\Lambda^{-1})_{b'}^{~a}) $.\\
\end{tabular}\right.
\end{eqnarray}
and the translation $T$ by
the amount of coordinate vector $y^a$ as
\begin{eqnarray} \label{T}
T(y^a)  &:&\left\{\begin{tabular}{clll}
    $ x^a$ & $ \to $& $x^{a'} ;$ & $ {x}^{a'}= x^a+ y^a$,\\
    $ e^a$ & $\rightarrow $& $e^{a'}$; &$ e^{a'} = e^a$, \\
   $ f$& $\rightarrow $&$ f'$ ; &$ f'(x^{a'}) = f(x^a)= f(x^{a'}- y^a) $. \\
\end{tabular}\right.
 \end{eqnarray}
Then, the infinitesimal transformation is given
in terms of $igl(4,R)$ generators:
\begin{eqnarray}\label{action}
\delta_\epsilon S= -i \epsilon^c Y_c \triangleright S.
\end{eqnarray}
The action of $M^a_{~b}$ is represented by
\begin{eqnarray} \label{M:x}
M^a_{~b}\triangleright x^c &=&-i x^a \delta_b^c, \quad
M^a_{~b}\triangleright e^c =-i e^a\delta_b^c, \\
\left(M^a_{~b}\triangleright f\right)(x^a) &=& -i x^a \frac{\partial}{\partial x^b} f(x^a)\,,\nn
\end{eqnarray}
and of $P_a$ by
\begin{eqnarray}\label{PA}
&& P_a \triangleright x^b=-i \delta_a^b, \quad
P_a \triangleright e^b = 0, \\
&&\left(P_a\triangleright f\right)(x^b)
= -i \frac{\partial}{\partial x^a} f(x^b) \,. \nn
\end{eqnarray}

Note that the translation and thus, the energy operator $E$
does not change the dual basis vector $e^a$.
On the other hand, the spatial dilatation operator $D$
non-trivially acts as:
\begin{eqnarray*}
&& \exp(i\alpha D)\triangleright x^a
  =  x_{(\alpha)}^a  \exp(i\alpha D)
    \triangleright\,, \quad
    \exp(i\alpha D) \triangleright e^a =
    e_{(\alpha)}^a\, \exp(i\alpha D) \triangleright\,, \\
&& (\exp(i\alpha D)\triangleright f)(x^a)
= f(x_{(-\alpha)}^a) ,
 \end{eqnarray*}
 where $ x_{(\alpha)}^a  =(x^0, \, \exp(\alpha)\, x^i)$
 and $ e_{(\alpha)}^a  =(e^0, \, \exp(\alpha)\, e^i)$.
This non-trivial transformation law provides
the $\ast$-product between the space coordinates
and/or the dual-basis vectors $\{ x^a,e^a\}$.
Between the two basis vectors, we have
\[e^a \ast e^b = \cdot [{\cal F}_{\kappa}^{-1} (e^a \otimes e^b)]=e^a e^b\,,
\] 
where the time translational invariance~(\ref{PA})
 $E \triangleright e^{a}=0$ is used.
Between  $e^a$ and $x^b$ we have
\begin{eqnarray}\label{A.e}
e^a \ast x^b &=& m [{\cal F}_{\kappa}^{-1} (e^a \otimes x^b)]
    =e^a x^b-\frac{i}{2\kappa} \delta^a_i \delta^b_0 e^i \,,
\nn\\
x^b \ast e^a &=& m [{\cal F}_{\kappa}^{-1} ( x^b \otimes e^a)]
    = e^a x^b+\frac{i}{2\kappa} \delta^a_i \delta^b_0 e^i \,,
\nn
\end{eqnarray}
which results in the commutation relation
\begin{eqnarray} \label{[x,e]}
[e^a,e^b]_\kappa=0,\quad [x^0, e^i]_\kappa = \frac{i}{\kappa} e^i,\quad [x^0, e^0]_\kappa=0=[x^i,e^a]_\kappa\,.
\end{eqnarray}

The twist deformation is also applied to
the multiplication of
two functions $f$ and $g$ which transform
according to~(\ref{Lambda:f}):
\begin{eqnarray}
 \label{kmult}
f(x)\ast g(x) \equiv m_\kappa[f\otimes g](x)&:=& m\left[\cFk^{-1}
    (f\otimes g)\right](x),
\end{eqnarray}
when the commutative multiplication is defined as $
m[f\otimes g](x) := f(x) g(x) $.
This twist deformation leads to
the conventional $\kappa$-Minkowski star product,
\begin{eqnarray}
 \label{kmoyal}
f(x)\ast g(x)&:=&
\left. \exp \left[ \frac{i}{2\kappa}
\Big(\fr{\del}{\del x_0} y^k\fr{\del}{\del y_k}
-x^k \fr{\del}{\del x_k}\fr{\del}{\del y_0} \Big) \right]
f(x)g(y)\right|_{x=y}.
\end{eqnarray}
Explicitly, the star product of two exponential functions
is given by
\begin{eqnarray}
e^{i px}\ast e^{iq x}
&=& m \left(\exp\left[-\frac{i}{2\kappa}
\left( E\otimes D-D\otimes E\right)\right]
\triangleright  (e^{i px}\otimes e^{iq x})\right) \\
  &=& m \left(\exp\left[-\frac{i}{2\kappa}
\left( p_0\otimes D-D\otimes q_0\right)\right]
\triangleright (e^{i px}\otimes e^{iq x})\right)
\nn\\
&=& e^{i (p_0+q_0)x^0+(p_ie^{\frac{q_0}{2\kappa}}
+q_ie^{-\frac{p_0}{2\kappa}})  x^i }\,.
\nn
\end{eqnarray}
Note that this twist deformation reproduces
the symmetric ordering result of the conventional
$\kappa$-Minkowski spacetime~\cite{sym:ordering}.
It should be noted that the action of $E$ and $D$
applies on $x$-space only and not on $p$ and $q$
which are just numbers.

On the star-product, the action $\triangleright_\kappa$
of $Y \in \{P_a,~M^a_{~b}\}$
is defined by
\begin{eqnarray}
\label{actiononthestar}
Y\triangleright_\kappa (A\ast B) &=&  (Y_{(1)}\triangleright_\kappa A)\ast  (Y_{(2)}\triangleright_\kappa B) \,,
\end{eqnarray}
where $Y_{(1,2)}$ is defined
in (\ref{def-coproduct}).
The action  $\triangleright_\kappa$
on the right-hand side reduces to
the undeformed one $\triangleright$
when $A$ or $B$ contains no $\ast$-product.
In addition, the $\kappa$ in $\triangleright_\kappa$ only implies that the action is acting on the star product. In general, we may omit the $\kappa$.

We now elaborate on the differential calculus
of the $\kappa$-Minkowski spacetime
based on the twist deformation.
The differential structure
constructed in Ref.~\cite{kappa-diff}
are five dimensional,
which was based on Jacobi identity and
the usual Leibnitz rule.
Here we will {\it not require the usual} Leibnitz rule.
Instead, we identify the partial derivative
with the generator element $i P_a$.
In this procedure,
Leibnitz rule is naturally modified
by the twist operation
and four dimensional differential calculus is obtained.

The derivative $ \partial_a= i P_a$
on the $\ast$-products is
defined according to (\ref{actiononthestar}) with
$\Delta_\kappa (\partial_a) = i \Delta_\kappa (P_a ) $
and is governed by the rule,
\begin{eqnarray}
\partial_0 \triangleright_\kappa(\phi(x)\ast \psi(x))
&=& (\partial_0 \triangleright_\kappa\phi(x))\ast \psi(x)
+ \phi(x)\ast (\partial_0 \triangleright_\kappa\psi(x)),
\label{Leibnitz} \\
\partial_i \triangleright_\kappa(\phi(x)\ast \psi(x))
&=&
(\partial_i \triangleright_\kappa\phi(x))\ast
\Big(e^{\frac E{2\kappa} }\psi(x)\Big)
+ \Big(e^{-\frac E{2\kappa} }\phi(x)\Big)
\ast (\partial_i \triangleright_\kappa\psi(x)) \,,
\nn
\end{eqnarray}
which gives the modified  Leibnitz rule
and is different from the one in~\cite{kappa-diff}.

Our definition of the derivative (\ref{Leibnitz})
has following nice properties.
First, the derivative rule is consistent
with the $\ast$-product.
Suppose one acts the derivative on
the exponential functions  $e^{i p x}$ in two ways.
One using the translation rule (\ref{PA}):
$P_a\triangleright e^{i p x} = p_a e^{i p x}$.
The other is to use the symmetric ordering
$ [e^{i p x}]_s  \equiv e^{i p_0 x^0/2}\ast e^{i \vec{p}\cdot \vec{x}}\ast e^{i p_0 x^0/2} $
and use the modified Leibnitz rule (\ref{Leibnitz}):
$P_a\triangleright_\kappa [e^{i p x}]_s
= P_a \triangleright_\kappa (e^{i p_0 x^0/2}
\ast e^{i \vec{p}\cdot \vec{x}}\ast e^{i p_0 x^0/2})
= p_a  [e^{i p x}]_s $.
For the action of the generator $M^a_{~b}$, we also have $M^a_{~b}\triangleright e^{i p x} = p_b x^a e^{i p x}$ and $M^a_b\triangleright_\kappa [e^{i p x}]_s
= p_b \triangleright_\kappa (e^{i p_0 x^0/2}
\ast (x^a e^{i \vec{p}\cdot \vec{x}})\ast e^{i p_0 x^0/2})
$.

Second, one can define the differential operator
$d_\kappa$ acting on a function $f$ or a $n$-form
$\omega= dx^{a_1}\wedge dx^{a_2}
\cdots\wedge dx^{a_n} f_{a_1 a_2\cdots a_n}$
\begin{eqnarray} \label{df}
&& d_\kappa f =  dx^a \ast (i P_a \triangleright_\kappa f)\,,\\
&& d_\kappa\omega =  dx^c\wedge dx^{a_1}\wedge dx^{a_2}\cdots\wedge dx^{a_n} \ast (i P_c\triangleright_\kappa f_{a_1 a_2\cdots a_n}) ,\nn\\
&&d_\kappa f \wedge d_\kappa g=-d_\kappa g \wedge d_\kappa f . \nn
\end{eqnarray}
In this calculation, we implicitly assume
that the wedge include the $\ast$-product.
(See the comment below (\ref{d2}).)
The definition of $d_\kappa$
allows differential calculus on the $\ast$-product.

Explicitly, one has
\begin{eqnarray}
\label{d-operation}
d_\kappa (x^0\ast x^i) &=& i dx^a \ast [P_a \triangleright_\kappa (x^0 \ast x^i)]
  = dx^0  \ast x^i + dx^j\ast  (e^{-E/(2\kappa)} x^0 \ast i (P_j \triangleright x^i))\\
   &=&(dx^0)  \ast x^i + (dx^i)\ast  x^0 + \frac{i}{2\kappa} dx^i \,,\nn \\
d _\kappa(x^i\ast x^0) &=& i dx^a \ast [P_a \triangleright_\kappa (x^i \ast x^0)]
  = dx^0 \ast x^i + dx^j\ast  ( i (P_j \triangleright x^i)\ast e^{E/(2\kappa)} x^0) \nn\\
   &=&(dx^0) \ast x^i + (dx^i) \ast x^0 -\frac{i}{2\kappa} dx^i . \nn
\end{eqnarray}
On the other hand, in the differential geometry,
the coordinate representation of $e^a$ is given by $dx^a$.
Therefore, from  Eq.~(\ref{[x,e]}),
we have commutation relations with differential elements
\begin{equation}
\label{[dx,dx]}
~[dx^a, dx^b]_\kappa=0,\quad
[x^0,dx^i]_\kappa= \frac{i}{\kappa} dx^i,
\quad[x^0, dx^0]_\kappa=0= [x^i, dx^a]_\kappa
.
\end{equation}
This relations are consistent with $d_\kappa$ operation:
\[d_\kappa[x^0,x^i]_\kappa= \frac{i}{\kappa}dx^i= [dx^0,x^i]_\kappa+[x^0,dx^i]_\kappa
\]
from (\ref{d-operation}) even though
the Leibnitz rule (\ref{Leibnitz}) in general shows $d_\kappa[f,g]_\kappa\neq [df,g]_\kappa+[f,dg]_\kappa$.

Third, one can also check that the general linear transformation acting on both side of Eq.~(\ref{[dx,dx]}) is consistent with the coproduct (\ref{iglcoproduct}).
Especially,
\begin{eqnarray}\label{M0l}
M^0_{~l}\triangleright_\kappa [x^i, dx^j]_\kappa = M^0_{~l}\triangleright_\kappa ( x^i \ast dx^j - dx^j \ast x^i) = 0,
\end{eqnarray}
which is ensured by the equations $[x^0, dx^j]_\kappa = \frac{i}{\kappa} dx^j$ and $[x^i, dx^0]_\kappa=0$, and the second equation in~(\ref{M:x}).
This result is also consistent with $M^a_{~b} \triangleright_\kappa (dx^c) = d(M^a_{~b}\triangleright_\kappa x^c)$, which comes from the postulate that the action of the general linear algebra extends to the differential algebra in a natural covariant way~\cite{kappa-diff}. 
The relations~(\ref{[dx,dx]}) has been also proposed in~\cite{Oeckl,amelino}, however the action of 
$\kappa$-Poincar\'e boost generators show that the relation~(\ref{M0l}) is not satisfied~\cite{kappa-diff}. 

Finally, the last equation of~(\ref{df}) and the commutativity of
$P_a\triangleright_\kappa$ operations ($P_a\triangleright_\kappa P_b\triangleright_\kappa = P_b\triangleright_\kappa P_a\triangleright_\kappa$) ensure the identity
\begin{equation}
\label{d2}
d_\kappa^2\omega =0
\end{equation}
and makes the $\ast$-product with
the wedge in~(\ref{df}) be irrelevant.
One may also confirm that the Jacobi identity such as
\[
[x^a, [x^b, dx^c]_\kappa]_\kappa + [x^b,[dx^c,x^a]_\kappa]_\kappa+[dx^c, [x^a, x^b]_\kappa]_\kappa=0
\]
is satisfied with the present differential structure~(\ref{[dx,dx]}).

In conclusion, 
our proposal provides the $\kappa$-covariance of the differential calculus~(\ref{[dx,dx]})
without introducing an extra dimensional differential. 
We have enlarged the carrier algebra of the twist function (from Poincar\'e to IGL(4,R)) but we preserve the 
classical algebra. 
This is in contrast with the proposal in~\cite{Ballest} 
which is based on much smaller extension of Poincar\'e algebra (from Poincar\'e to Weyl) but the classical algebra is
also modified.
It will be interesting to find the differential structure
explicitly in this approach.

\begin{acknowledgments}
We thank J. Lukierski for introducing related references and for useful comments about this paper.
This work was supported by SRC Program of the KOSEF through the CQUEST grant R11-2005-021 (K\&R)
and by the Korea Research Foundation Grant funded by the Korean Government(MOEHRD)(KRF-2007-355-C00013(L); KRF-2007-313-C00153(R)).
\end{acknowledgments} 

\vspace{4cm}

\end{document}